\documentclass[a4paper,11pt]{article}

\usepackage{soul}
\usepackage{color}
\usepackage[singlelinecheck=off,center]{caption}
\usepackage[T2A]{fontenc}
\usepackage[utf8x]{inputenc}
\usepackage[warn]{mathtext}
\usepackage{indentfirst}
\usepackage[fleqn]{amsmath}
\usepackage{amsthm,amsfonts,amssymb,amscd}
\usepackage{hyperref}
\usepackage{bm}
\usepackage{icomma}
\usepackage{mathrsfs}

\renewcommand{\vec}{\mathbf}

\begin{document}

\title{Continuous measurements in probability representation of quantum mechanics}
\author{Y. V. Przhiyalkovskiy \\ \\ \textit{\small Kotelnikov Institute of Radioengineering and Electronics (Fryazino Branch)}\\ \textit{\small of Russian Academy of Sciences,}\\ \textit{\small Vvedenskogo sq., 1, Fryazino, Moscow reg., 141190, Russia} \\ \small e-mail: yankus.p@gmail.com}
\date{}

\maketitle

\begin{abstract}
The continuous quantum measurement within the probability representation of quantum mechanics is discussed. The partial classical propagator of the symplectic tomogram associated to a particular measurement outcome is introduced, for which the representation of a continuous measurement through the restricted path integral is applied. The classical propagator for the system undergoing a non-selective measurement is derived by summing these partial propagators over the entire outcome set. The elaborated approach is illustrated by considering non-selective position measurement of a quantum oscillator and a particle.
\end{abstract}

\section{Introduction}

The probability representation of quantum mechanics, introduced not so far and being actively developed in recent years, is attractive due to being one of the most promising to formulate quantum mechanics in the closest manner to statistical physics \cite{Chernega2020, Chernega2019, Korennoy2017, MManko2013}. Essentially, this approach suggests that a family of probability distributions of a coordinate in linearly and homogeneously transformed phase space is employed to describe a quantum state instead of a density matrix. Due to their unambiguous mapping to each other, it is turns out to be possible to formulate quantum mechanics in terms of such probability distributions, or so-called quantum tomograms. 

The importance of influence the measuring of an observable exerts on a quantum system could hardly be underestimated from both a theoretical and practical viewpoint. How a measuring process is reflected within the probability representation is also of great interest. According to the original quantum theory, the measurement of an observable performed on the system happens instantaneously and thus implies the collapse of its state. The same obviously applies to the quantum tomogram. In real systems, instead, the state the system had before the measurement transits to the new state in a continuous way during the measurement. The profound research of this subject in quantum mechanics has begun in 70-th \cite{Davies1969, Davies1970, Davies1971, Davies1970-2} and is well studied now. In practice, the issues of decoherence and measurement back-action comprise a significant part of current research on quantum computing, actively growing in our days \cite{Pellizzari1995, Beige2000, Albash2015}. Moreover, apart from being traditionally considered as a passive, continuous measurement may even be involved to manipulate a quantum system \cite{Pechen2006, Shuang2007, Shuang2008, Pechen2015}. Thus, measuring of a predetermined set of observables ensures the final state of the particular system to have maximum expected value of a target operator \cite{Pechen2006}. Another promising application of active measuring is the optimal control of quantum evolution, in particular, employing quantum Zeno and anti-Zeno effects \cite{Pechen2006, Shuang2007, Shuang2008} or optimal acceleration of the Landau-Zener transitions \cite{Pechen2015}. The most intuitively clear concept to describe a continuous measurement of a quantum system is based on the Feynman path integral and was elaborated by Mensky \cite{Mensky1979, Mensky1993}. The main idea of this approach is that certain paths over which the integral is taken are more preferred, according to the information the environment gains from the measured system. Technically, it is attained by inserting a path weight functional into the path integral to calculate the quantum propagator. 

Certainly, the continuous measurement experienced by the system modifies both the tomogram dynamics and the resultant tomogram. A direct parallel drawn between traditional quantum mechanics and its probability representation leads to a differential Fokker-Plank-type equation whose solution determines the tomogram for every time moment \cite{Mancini1997}. The other approach to figure out the evolved tomogram, elaborated so far only for isolated systems, is to use a so-called classical propagator \cite{Manko1998, OManko1999, Fedorov2013}. Being attributed to a particular quantum system, the classical propagator is determined by the usual quantum propagator for that system and hence already incorporates its dynamics. The central theme we focus our attention in this article on is expansion of the approach of classical propagators in symplectic tomography to quantum systems undergoing continuous measurement by application the restricted path integral.

\section{Symplectic tomography}

Symplectic tomography was initially introduced in \cite{Mancini1997, Mancini1995, Mancini1996, OManko1997} in the following way. Consider a classical system with its phase space, and let an observable $X = \mu q + \nu p$ be the result of a general linear transformation of the coordinate $q$ and momentum $p$ (hereafter $\hbar = 1$). In other words, $X$ is the coordinate in a phase space viewed from a new frame according to the above transform. Here the real quantities $\mu$ and $\nu$ parametrize this map, after performing of which the coordinate $X$ is measured. If we now turn to the quantum counterpart of the system, its symplectic tomogram is then, by definition, the marginal distribution function of the variable $X$. Namely, it is the Fourier transformation of the characteristic function $F(k) = \langle e^{i k \hat{X}} \rangle$ related to the self-adjoint operator $\hat{X} = \mu \hat{q} + \nu \hat{p}$:
\begin{equation}
	\mathcal{T}(X, \mu, \nu) = \frac{1}{2 \pi} \int \langle e^{i k \hat{X}} \rangle e^{-i k X} dk
	\label{eq:T_definition}
\end{equation}
where $\langle \hat{A} \rangle = \operatorname{Tr} \{\hat{\rho} \hat{A} \}$ is the quantum mean value. If one performs the subsequent reparametrization $\mu = \cos{\theta} e^\lambda$ and $\nu = \sin{\theta} e^{-\lambda}$, it becomes clear the physical sense of the above transformation of the phase space as the rotation and scaling \cite{Ibort2009}. In particular, for $\lambda = 1$ the map reduces to Radon transformation, and the symplectic tomogram turns out to be an optical tomogram \cite{Lvovsky2009}.

The symplectic tomogram defined in this way is positive definite and satisfies 
\begin{equation}
	\int \mathcal{T}(X, \mu, \nu) dX = 1.
\end{equation}
Therefore, the tomogram $\mathcal{T}(X, \mu, \nu)$ indeed turns out to be a probability distribution for each $\mu$ and $\nu$. The essential feature of such a tomogram set is that it is equivalent to a quantum state. Consequently, it is possible to formulate quantum mechanics taking this set of probability distributions as a system state. It is just this formulation that has been called the probability representation of quantum mechanics.

\subsection{Star-product formalism}

To begin with, we review the framework of operator symbols \cite{OManko2009}. Essentially, its purpose is to relate the algebras of Hilbert space operators with algebras of ordinary functions equipped with an associative but non-commutative product, the so-called star product.

Consider a Hilbert space $\mathbb{H}$ attributed to the quantum system and an operator $\hat{A}$ acting on vectors of $\mathbb{H}$. Let $\vec{x}_i = (x_i^1, x_i^2, \dots, x_i^k)$ be a vector of parameters. Let also $\hat{\mathcal{U}}(\vec{x})$ and $\hat{\mathcal{D}}(\vec{x})$ be operators parameterized by $\vec{x}_i$ and satisfying the consistency condition $\operatorname{Tr} \{ \hat{\mathcal{U}}(\vec{x}_1) \hat{\mathcal{D}}(\vec{x}_2) \} = \delta(\vec{x}_1 - \vec{x}_2)$. Then, one defines the transformation of the operator $\hat{A}$ into a C-function $f_{\hat{A}}(\vec{x})$ through 
\begin{equation}
	f_{\hat{A}}(\vec{x}) = \operatorname{Tr}\{ \hat{A} \hat{\mathcal{U}}(\vec{x}) \}
	\label{eq:symbol_through_A}
\end{equation}
and the inverse transformation as
\begin{equation}
	\hat{A} = \int f_{\hat{A}}(\vec{x}) \hat{\mathcal{D}}(\vec{x}) d\vec{x}
	\label{eq:A_through_symbol}
\end{equation}
where $d\vec{x} = dx^1 dx^2 \dots dx^k$. The operators $\hat{\mathcal{U}}(\vec{x})$ and $\hat{\mathcal{D}}(\vec{x})$ are referred to as the dequantizer and quantizer respectively, and the function $f_{\hat{A}}(\vec{x})$ is then the operator symbol of $\hat{A}$. 

The symbol of a multiplication of two operators, $\hat{A} \hat{B}$, can be easily derived using \eqref{eq:symbol_through_A} and \eqref{eq:A_through_symbol}:
\begin{equation}
	f_{\hat{A} \hat{B}}(\vec{x}) = \iint f_{\hat{A}}(\vec{x}_1) f_{\hat{B}}(\vec{x}_2) M_{\vec{x}_1\vec{x}_2}(\vec{x}) d\vec{x}_1 d\vec{x}_2
	\label{eq:ABproduct} 
\end{equation}
where the kernel is
\begin{equation}
	M_{\vec{x}_1\vec{x}_2}(\vec{x}) = \operatorname{Tr} \left\{ \hat{\mathcal{U}}(\vec{x}) \hat{\mathcal{D}}(\vec{x}_1) \hat{\mathcal{D}}(\vec{x}_2) \right\}.
\end{equation}
Relation \eqref{eq:ABproduct}, which is also convenient to shortly denote by
\begin{equation}
	f_{\hat{A} \hat{B}}(\vec{x}) = f_{\hat{A}}(\vec{x}) \star f_{\hat{B}}(\vec{x}),
	\label{eq:star_product}
\end{equation}
is called the star product of operator symbols. Using \eqref{eq:ABproduct}, one obtains the symbol of a commutator
\begin{equation}
	f_{[\hat{A}, \hat{B}]}(\vec{x}) = \iint f_{\hat{A}}(\vec{x}_1) f_{\hat{B}}(\vec{x}_2) C_{\vec{x}_1 \vec{x}_2}(\vec{x}) d\vec{x}_1 d\vec{x}_2
\end{equation}
with the kernel
\begin{equation}
	C_{\vec{x}_1 \vec{x}_2}(\vec{x}) = \operatorname{Tr} \left\{ \hat{\mathcal{U}}(\vec{x}) [\hat{\mathcal{D}}(\vec{x}_1), \hat{\mathcal{D}}(\vec{x}_2)] \right\} = M_{\vec{x}_1 \vec{x}_2}(\vec{x}) - M_{\vec{x}_2 \vec{x}_1}(\vec{x}).
	\label{eq:Commutator_kernel}
\end{equation}
Similarly, a commutator symbol can be written in a compact manner, just as
\begin{equation}
	f_{[\hat{A}, \hat{B}]}(\vec{x}) = [f_{\hat{A}}(\vec{x}), f_{\hat{B}}(\vec{x})]_\star
\end{equation}
where $[~, ~]_\star$ denotes a star-product commutator: $[f, g]_\star = f \star g - g \star f$.

\subsection{Symplectic tomography using operator symbols}

The symplectic tomography can be conveniently introduced using the framework of operator symbols \cite{OManko2009}, briefly reviewed in the preceding section. Specifically, let the parameters set be $\vec{x} = (X, \mu, \nu)$ and then define the dequantizer and quantizer operators as
\begin{subequations}
	\begin{align}
		& \hat{\mathcal{U}}(\vec{x}) = \delta(X - \mu \hat{q} - \nu \hat{p}), \label{eq:U} \\
		& \hat{\mathcal{D}}(\vec{x}) = \frac{1}{2\pi} \exp{\left( iX - i \mu \hat{q} - i\nu \hat{p} \right)}. \label{eq:D}
	\end{align}
	\label{eq:UD}
\end{subequations}
Here $\delta(x)$ denotes the Dirac delta function, in the case of an operator argument being treated as $\delta{(\hat{A})} = (2\pi)^{-1} \int dk e^{i k \hat{A}}$. A symplectic tomogram is, by definition, a symbol of the density matrix calculated using above dequantizer operator \eqref{eq:U}:
\begin{equation}
	\mathcal{T}(\vec{x}) = \operatorname{Tr} \{ \hat{\rho} \hat{\mathcal{U}}(\vec{x}) \}.
	\label{eq:Tomogram_definition}
\end{equation}
A simple comparison reveals the equivalence of this definition to \eqref{eq:T_definition}. Having the tomogram, the density matrix is easily restored by the inverse transformation using quantizer \eqref{eq:D}:
\begin{equation}
	\hat{\rho} = \int \mathcal{T}(\vec{x}) \hat{\mathcal{D}}(\vec{x}) d\vec{x}.
	\label{eq:inverse_transformation}
\end{equation}
By a direct calculation, one also derives an explicit expression of the operator multiplication kernel
\begin{equation}
	M_{\vec{x}_1 \vec{x}_2} (\vec{x}) = \frac{\delta(\mu (\nu_1 + \nu_2) - \nu (\mu_1 + \mu_2))}{4\pi^2} e^{iX_1 + iX_2} e^{-i \frac{(\nu_1 + \nu_2)}{\nu} X} e^{ i (\nu_1\mu_2 - \nu_2\mu_1) /2}
\end{equation}
being needed to carry out the further calculations.

\subsection{Symplectic tomogram evolution}

Consider a quantum system which dynamics is described by a certain Hamiltonian $\hat{H}$. The evolution of the system state, being mixed in general, is described by the evolution operator $\hat{U}_{t} = \exp{( -i \widehat{H} t )}$. The density matrix of the system at time $t>0$ is then expressed through the initial density matrix at time $t=0$ as
\begin{equation}
	\hat{\rho}(t) = \hat{U}_{t} \hat{\rho}(0) (\hat{U}_{t})^\dag.
\end{equation}
Each matrix element $U_{t}(q_{f}, q_{i}) \equiv \langle q_{f} | \hat{U}_{t} | q_{i} \rangle$ of the evolution operator, which is the amplitude of the system transition from the point $q_{i}$ at time $0$ to the point $q_{f}$ at time $t$, as known, can be expressed through the Feynman path integral \cite{Feynman2010}
\begin{equation}
	U_{t}(q_{f}, q_{i}) = \int\limits_{q_{i}, 0}^{q_{f}, t} d[q] e^{i S[q]}
	\label{eq:evolution_operator}
\end{equation}
where $S$ is the action of the system.

It is clear, that at each instant the tomogram of the quantum system can be expressed through the density matrix at that time,
\begin{equation}
	\mathcal{T}(\vec{x}, t) = \operatorname{Tr} \{ \hat{\rho}(t) \hat{\mathcal{U}} \}.
\end{equation}
This equation means that the tomogram is the instant average value of the dequantizer $\hat{\mathcal{U}}$ eigenvalues. Turning to the Heisenberg picture, the tomogram 
\begin{equation}
	\mathcal{T}(\vec{x}, t) = \operatorname{Tr} \{ \hat{\rho}(0) \hat{\mathcal{U}}_{t} \}
\end{equation}
makes sense of the eigenvalues average of the operator $\hat{\mathcal{U}_t} \equiv (\hat{U}_{t})^\dag \hat{\mathcal{U}} \hat{U}_{t}$, calculated using the initial density matrix. Hereafter, the latter operator will be referred to as evolved dequantizer. 

Applying the inverse transformation \eqref{eq:inverse_transformation} to initial density matrix $\hat{\rho}(0) = \int \mathcal{T}(\vec{x}, 0) \hat{\mathcal{D}}(\vec{x}) d\vec{x}$, one can relate the evolved tomogram with the initial one by
\begin{equation}
	\mathcal{T}(\vec{x}, t) = \int \mathcal{T}(\vec{x}', 0) \Pi_t(\vec{x}', \vec{x}) d\vec{x}'
\end{equation}
where 
\begin{equation}
	\Pi_t(\vec{x}', \vec{x}) = \operatorname{Tr} \left\{  \hat{\mathcal{D}}(\vec{x}') \hat{\mathcal{U}}_{t}(\vec{x}) \right\} = \operatorname{Tr} \left\{  \hat{\mathcal{D}}(\vec{x}') (\hat{U}_{t})^\dag \hat{\mathcal{U}}(\vec{x}) \hat{U}_{t} \right\}
	\label{eq:tom_propagator_w_o_measurement}
\end{equation}
is referred to as the "classical" propagator (or tomogram propagator), in contrast to quantum propagator \eqref{eq:evolution_operator} \cite{OManko1999}. Substituting then dequantizer and quantizer operators \eqref{eq:UD} into \eqref{eq:tom_propagator_w_o_measurement} yields the explicit expression
\begin{equation}
	\begin{aligned}
		& \Pi_t(\vec{x}', \vec{x}) = \\
		& = \frac{1}{4\pi^2} \int k^2 U_t^*(q_{f,1} + k\nu, q_{i,2}) U_t(q_{f,1}, q_{i,2} + k\nu') e^{ ik(X' + X)} e^{-i k^2 \frac{\mu' \nu' + \mu \nu}{2}} e^{- i k\mu' q_{i,2} - ik\mu q_{f,1}} dq_{f,1} dq_{i,2} dk \\
	\end{aligned}
	\label{eq:isolated_classical_propagator}
\end{equation}
where the propagator homogeneity of degree $-2$ relative to the first argument, $\Pi_t(k\vec{x}', \vec{x}) = k^{-2} \Pi_t(\vec{x}', \vec{x})$, has been used, directly stemming from the absolute homogeneity of the tomogram $\mathcal{T}(k\vec{x}) = |k|^{-1} \mathcal{T}(\vec{x})$.

\section{Continuous measurements in symplectic tomography}

As long as we consider an isolated quantum system, all paths in the configuration space connecting the starting and end points must be involved when Feynman path integral \eqref{eq:evolution_operator} is calculated. A continuous measuring of the system, however, implies that a certain information about the system is acquired by environment. Therefore, we have some knowledge about the path along which the system has passed. This can be quantified by introducing a path weight functional $w_a[q]$, $0 \le w_a[q] \le 1$, where $a(t)$ is the measurement outcome. In other words, all paths we integrate over are weighted by $w_a[q]$ according to the probability of passing through. Therefore, the Feynman path integral used to get the transition amplitude is generalized to \cite{Mensky1994}
\begin{equation}
	U_{t, a}(q_{f}, q_{i}) = \int\limits_{q_{i}, 0}^{q_{f}, t} d[q] w_a[q] e^{i S[q]}.
	\label{eq:weighted_evolution_operator}
\end{equation}
It is important to note that the transition amplitude defined in this way is actually no longer unitary due to the path weighting. Given that the measurement result is $a(t)$, the system state after the measurement is then described by density matrix
\begin{equation}
	\hat{\rho}_a(t) = \hat{U}_{t, a} \hat{\rho}(0) (\hat{U}_{t, a})^\dagger
	\label{eq:state_a}
\end{equation}
which is obviously not normalized. Therefore, the tomogram density in the set of outcomes $\{ a \}$ is
\begin{equation}
	\mathcal{T}_a(\vec{x}, t) = \operatorname{Tr} \{ \hat{\rho}_a(t) \hat{\mathcal{U}} \} = \operatorname{Tr} \{ \hat{\rho}(0) \hat{\mathcal{U}}_{t, a} \}
\end{equation}
where $\hat{\mathcal{U}}_{t, a} \equiv (\hat{U}_{t, a})^\dagger \hat{\mathcal{U}} \hat{U}_{t, a}$ is the evolved dequantizer operator related to the measurement outcome $a(t)$. It is important to emphasize that $\mathcal{T}_a(\vec{x}, t)$ is not a true tomogram, since it was derived using a non-normalized density matrix. Although, it can become such were it normalized by either the outcome probability (the discrete spectrum case) or the probability the outcome lies in a certain range (the continuous spectrum case).

The tomogram density of the system undergone the selective measurement is related to the initial tomogram $\mathcal{T}(\vec{x}, 0)$ by
\begin{equation}
	\mathcal{T}_a(\vec{x}, t) = \int \mathcal{T}(\vec{x}', 0) \Pi_{t,a}(\vec{x}', \vec{x}) d\vec{x}'
\end{equation}
where
\begin{equation}
	\begin{aligned}
		& \Pi_{t,a}(\vec{x}', \vec{x}) = \operatorname{Tr} \left\{ \hat{\mathcal{D}}(\vec{x}') (\hat{U}_{t,a})^\dag \hat{\mathcal{U}}(\vec{x}) \hat{U}_{t,a} \right\} = \\
		& = \frac{1}{4\pi^2} \int k^2 U_{t,a}^*(q_{f,1} + k\nu, q_{i,2}) U_{t,a}(q_{f,1}, q_{i,2} + k \nu') e^{ ik(X' + X)} e^{-i k^2 \frac{\mu' \nu' + \mu \nu}{2}} e^{- i k\mu' q_{i,2} - ik\mu q_{f,1}} dq_{f,1} dq_{i,2} dk
	\end{aligned}
	\label{eq:selective_propagator}
\end{equation}
is the partial propagator. Note that in contrast to propagator \eqref{eq:isolated_classical_propagator}, the partial propagator maps the initial tomogram to the tomogram density.

If we consider non-selective measurement, the result is presumed to be unknown. Therefore, to obtain the final density matrix of the system, one must integrate density matrices \eqref{eq:state_a} over all states relevant to particular outcomes \cite{Mensky1993},
\begin{equation}
	\hat{\rho}(t) = \int \hat{U}_{t, a} \hat{\rho}(0) (\hat{U}_{t, a})^\dagger da,
	\label{eq:NSM_density_matrix}
\end{equation}
where $d a$ is the measure in the set of outcomes. Note here, that from \eqref{eq:NSM_density_matrix} it immediately follows that the generalized unitarity condition 
\begin{equation}
	\int (\hat{U}_{t, a})^\dagger \hat{U}_{t, a} da = 1
\end{equation}
must be fulfilled to ensure $\operatorname{Tr} \{\hat{\rho}(t) \} = 1$. The tomogram of the system undergone by a non-selective measurement is obtained by calculating the symbol of matrix \eqref{eq:NSM_density_matrix}, yielding
\begin{equation}
	\tilde{\mathcal{T}}(\vec{x}, t) = \int \mathcal{T}_a(\vec{x}, t) da = \operatorname{Tr} \{ \hat{\rho}(0) \hat{\tilde{\mathcal{U}}}_t \}
	\label{eq:T_measured}
\end{equation} 
where $\hat{\tilde{\mathcal{U}}}_t = \int (\hat{U}_{t, a})^\dagger \hat{\mathcal{U}} \hat{U}_{t, a} da$ is the evolved dequantizer for the measured system. 

In a similar way as for tomogram density, the application of inverse transformation \eqref{eq:inverse_transformation} to \eqref{eq:T_measured} yields the relation between the initial and final tomograms:
\begin{equation}
	\tilde{\mathcal{T}}(\vec{x}, t) = \int \mathcal{T}(\vec{x}', 0) \tilde{\Pi}_t(\vec{x}', \vec{x}) d\vec{x}'
	\label{eq:T_nonselective}
\end{equation}
where the tomogram propagator for the system under non-selective continuous measurement is just an integral of the partial propagators over the outcome set
\begin{equation}
	\tilde{\Pi}_t(\vec{x}', \vec{x}) = \int \Pi_{t,a}(\vec{x}', \vec{x}) da.
	\label{eq:non-selective_propagator}
\end{equation}
We especially stress here that since $\tilde{\Pi}_t(\vec{x}', \vec{x})$ maps tomograms, it is a true propagator, yet it contains the influence of the measuring environment.

In the conclusion consider the case of a non-selective continuous measurement of a single observable $\hat{A}$. In the traditional formulation of quantum mechanics, the system density matrix evolves according to the master equation of the form \cite{Mensky1994, Jacobs2006}
\begin{equation}
	\frac{\partial \hat{\rho}}{\partial t} = - i [\hat{H}, \hat{\rho}] - k [\hat{A}, [\hat{A}, \hat{\rho}]].
\end{equation}
Now turn to the probability representation. Replacing the operators by their symbol functions following \eqref{eq:symbol_through_A} and \eqref{eq:Tomogram_definition}, and subsequently using expansion \eqref{eq:T_nonselective}, one thus obtains that the propagator obeys
\begin{equation}
	\frac{\partial \tilde{\Pi}_t(\vec{x}', \vec{x})}{\partial t} + i [ f_{\hat{H}}(\vec{x}), \tilde{\Pi}_t(\vec{x}', \vec{x})]_\star + k [f_{\hat{A}}(\vec{x}), [f_{\hat{A}}(\vec{x}), \tilde{\Pi}_t(\vec{x}', \vec{x})]_\star]_\star = 0 
	\label{eq:propagator_master_equation}
\end{equation}
with initial condition $\tilde{\Pi}_0(\vec{x}', \vec{x}) = \delta \left( \vec{x}' - \vec{x} \right)$.

\section{Spectral measurement of oscillator position}

The harmonic oscillator is the underlying model being one of the most important in quantum mechanics. The basic investigations of an isolated oscillator in terms of probability representation of quantum mechanics have already been performed by now \cite{OManko1999}. Nevertheless, in real systems, the measuring environment will inevitably influence the oscillator, which requires revising its symplectic tomogram dynamics. In this section, we will demonstrate this for a driven quantum oscillator undergoing a continuous spectral measurement of its coordinate.

Assume the Hamiltonian of the oscillator to be $\hat{H} = \hat{p}^2/(2m) + m \omega^2 \hat{q}^2/2 - \hat{q} f(t)$ where $f(t)$ is the external force. The action used to calculate the path integral in \eqref{eq:weighted_evolution_operator} is then
\begin{equation}
	S = \int\limits_0^T dt \left( \frac{m}{2} \dot{q}^2 - \frac{m\omega^2}{2} q^2 + f q \right).
	\label{eq:action}
\end{equation}
Let the oscillator position be measured during the time interval $(0, T)$. To get further, decompose the oscillator trajectory $q(t)$ into the Fourier series as
\begin{equation}
	q(t) = \sum_{n = 1}^\infty q_n \sin{(\Omega_n t)}
\end{equation}
where $\Omega_n = \pi n/T$. According to the approach of the spectral measurement, one measures the component amplitudes $q_n$ with the errors $\Delta a_n$ \cite{Feynman2010}. The appropriate choice of a weight functional for such a measurement is \cite{Mensky1993, Menskii1988}
\begin{equation}
	w_{a} [q] = \exp \left\{ - \sum\limits_{n=1}^{\infty} \frac{(q_n - a_n)^2}{\Delta a_n^2} \right\}
	\label{eq:w_a_n}
\end{equation}
where $a_n$ is the measurement result for amplitude $q_n$. 

If the measurement result $\{a_n\}$ are known, the amplitude $U_{t, a}(q_{f}, q_{i})$ can be obtained by changing the integration over paths in path integral \eqref{eq:weighted_evolution_operator} to integration over their Fourier components. Substituting weighting functional \eqref{eq:w_a_n}, action \eqref{eq:action} (where the path $q(t)$ is expressed through its Fourier components) into \eqref{eq:weighted_evolution_operator} eventually yields
\begin{equation}
	U_{T, a}(q_{f}, q_{i}) = \left( \frac{1}{2\pi i} \prod\limits_{n=1}^{\infty} \left\{ 1 - \frac{\omega_{e,n}^2}{\Omega_n^2} \right\}^{-1} \right)^{1/2} \exp \left\{ i S(\eta) - \sum\limits_{n=1}^{\infty} \frac{(\eta_n - a_n)^2}{\Delta a_n^2} \frac{\Omega_n^2 - \omega^2}{\Omega_n^2 - \omega_{e,n}^2} \right\}
	\label{eq:oscillator_selective_U}
\end{equation}
where
\begin{equation}
	\omega_{e,n}^2 = \omega^2 - \frac{4 i}{m T \Delta a_n^2}
\end{equation}
and $\eta_n = (2/T) \int_0^T \eta(t) \sin{\Omega_n t} dt$ is the Fourier component amplitudes of the classical trajectory $\eta(t)$ derived from the equation of motion $d^2 \eta / d t^2 + \omega^2 \eta = f(t)/m$ for a classical oscillator with initial conditions $\eta(0) = q_{i}$, $\eta(T) = q_{f}$.

In the case of a non-selective coordinate measurement, the tomogram propagator is obtained by substituting partial quantum propagator \eqref{eq:oscillator_selective_U} into partial tomogram propagator \eqref{eq:selective_propagator} and subsequently integrating over the outcome set, following \eqref{eq:non-selective_propagator}. A direct but somewhat tedious calculation then results in
\begin{equation}
	\begin{aligned}
		& \tilde{\Pi}_T(\vec{x}', \vec{x}) = \\
		& = \left( \frac{1}{2 \pi \sigma^2} \right)^{1/2} e^{- \frac{\left( X - X' - \bar{X} \right)^2}{2 \sigma^2}} \delta \left(\mu' - \mu \cos{\omega T} +  \nu m \omega \sin{\omega T} \right) \delta \left( \nu' - \nu \cos{\omega T} - \mu \frac{\sin{\omega T}}{m \omega} \right).
	\end{aligned}
	\label{eq:oscillator_tilde_Pi}
\end{equation}
We see that a continuous coordinate measurement makes a dependence on $X'$ to be Gaussian with the variance
\begin{equation}
	\sigma^2 = 2 \kappa \left(\nu^2 (1 + \cos^2{\omega T}) + \mu \nu \frac{\sin{2\omega T}}{m \omega}  + \mu^2 \frac{\sin^2{\omega T}}{m^2 \omega^2} \right) - 4 \xi \left( \nu^2 \cos{\omega T} + \mu \nu \frac{\sin{\omega T}}{m \omega} \right)
	\label{eq:variance}
\end{equation}
and the shifted mean
\begin{equation}
	\bar{X} =  \int\limits_0^T f(t) \left( \nu \frac{\sin{\omega (T - t)} \cos{\omega T} + \sin{\omega t}}{\sin{\omega T}} + \mu \frac{\sin{\omega (T - t)}}{m \omega} \right) dt
\end{equation}
whenever the oscillator is acted by external force $f(t)$. The coefficients $\kappa$ and $\xi$ in \eqref{eq:variance} are determined by the measurement accuracy $\{\Delta a_n\}$ and amount to \cite{Menskii1988}
\begin{equation}
	\begin{aligned}
		& \kappa = \frac{2}{T^2} \sum\limits_{n=1}^{\infty} \frac{\Omega_n^2}{\Delta a_n^2 (\Omega_n^2 - \omega^2)^2}, \\
		& \xi = \frac{2}{T^2} \sum\limits_{n=1}^{\infty} \frac{(-1)^n \Omega_n^2}{\Delta a_n^2 (\Omega_n^2 - \omega^2)^2}. \\
	\end{aligned}
	\label{eq:precision_coefficients_n}
\end{equation}

Note that, given that propagator \eqref{eq:oscillator_tilde_Pi} of the measured oscillator has a Gaussian dependence on $X'$, integrating it with the initial tomogram in \eqref{eq:T_nonselective} blurs its dependence on $X$. Instead, in the limit $\Delta a_n \rightarrow \infty$ when there is actually no measurement, the variance \eqref{eq:variance} tends to zero, collapsing the Gaussian in propagator \eqref{eq:oscillator_tilde_Pi} to the $\delta$-function. The latter means that without the position measuring, the propagator does not change the dependence of the initial tomogram on $X$ at all.

\section{Particle scattering}

Another example, quite simple but worth discussing, is a particle scattering upon measuring its position. To consider the measurement as direct, we further assume $\Delta a_n = \Delta a$ for all $n$. Indeed, it is easy to show that in this case functional \eqref{eq:w_a_n} becomes
\begin{equation}
	w_{a} [q] = \exp \left\{ - \frac{2}{T \Delta a^2} \int\limits_0^T (q(t) - a(t))^2 dt \right\}.
\end{equation}

A particle can be obviously treated as the particular case of the oscillator having $\omega = 0$. Thus, taking the $\omega \rightarrow 0$ limit in propagator \eqref{eq:oscillator_tilde_Pi}, it reduces to
\begin{equation}
	\tilde{\Pi}_T(\vec{x}', \vec{x}) = \left( \frac{1}{2 \pi \sigma^2} \right)^{1/2} e^{- \frac{\left( X' - X - \bar{X} \right)^2}{2 \sigma^2}} \delta \left(\mu' - \mu \right) \delta \left( \nu' - \nu - \mu \frac{T}{m} \right)
	\label{eq:FP_propagator}
\end{equation}
becoming the propagator for a particle undergoing a position measurement. In this case, the variance $\sigma^2$ becomes
\begin{equation}
	\sigma^2 = \frac{2}{3 \Delta a^2} \left(3\nu^2 + 3 \nu \mu \frac{T}{m}  + \mu^2 \frac{T^2}{m^2}  \right)
	\label{eq:particle_propagator_variance}
\end{equation}
where it has been taken into account that coefficients \eqref{eq:precision_coefficients_n} containing the measurement accuracy are simplified to
\begin{equation}
	\kappa = - 2 \xi = \frac{1}{3 \Delta a^2}.
\end{equation}
Besides, the mean of $X'$, denoted above as $\bar{X}$, is now also simplified to
\begin{equation}
	\bar{X} = \int\limits_0^T f(t) \left( \nu + \mu \frac{T - t}{m} \right) dt.
\end{equation}

Let, for example, the particle has a position-space wave function
\begin{equation}
	\Psi_p(q) = (\pi l^2)^{-1/4} \exp \left\{ ipq - \frac{q^2}{2l^2} \right\}
\end{equation}
at initial time $t=0$. This state describes a particle located around the origin with Gaussian distribution with the deviation $\Delta q = l/\sqrt{2}$ and having an average impulse $p$ with uncertainty $\Delta p = 1/(\sqrt{2} l)$. The initial tomogram derived from \eqref{eq:Tomogram_definition} is then
\begin{equation}
	\mathcal{T}(\vec{x}, 0) = \left( \frac{\pi}{\mu^2 l^2 + \nu^2l^{-2} } \right)^{1/2} \exp \left\{ - \frac{\left( X - \nu p\right)^2}{\mu^2 l^2 + \nu^2l^{-2}} \right\}.
\end{equation}
Convolution of this tomogram with propagator \eqref{eq:FP_propagator} immediately gives the evolved particle tomogram at time $t=T$:
\begin{equation}
	\mathcal{T}(\vec{x}, T) = \left( \frac{\pi}{2 \sigma^2 + \mu^2 l^2 + \left( \nu + \mu \frac{T}{m} \right)^2 l^{-2}} \right)^{1/2} \exp \left\{ - \frac{\left( X + \bar{X} - \left( \nu + \frac{T}{m} \mu \right) p \right)^2}{2\sigma^2 + \mu^2 l^2 + \left( \nu + \mu \frac{T}{m} \right)^2 l^{-2}} \right\}
	\label{eq:T_particle_measured}
\end{equation}
where $\sigma^2$ is defined in \eqref{eq:particle_propagator_variance}. One sees that the continuous measurement of the particle position results, as expected, in an additive broadening of the initial distribution of $X$ by $\sigma^2$.

It is instructive to examine the change in the entropy of the particle, if its position has been measured. In general, the symplectic entropy reads \cite{Manko2006}
\begin{equation}
	S(\mu, \nu) = -\int \mathcal{T}(\vec{x}) \ln \mathcal{T}(\vec{x}) dX.
\end{equation}
For a Gaussian-type tomogram, as \eqref{eq:T_particle_measured} is, the direct calculation gives
\begin{equation}
	S_T(\mu, \nu) = \frac{1 + \ln \pi}{2} + \frac{1}{2} \ln \left[ 2\sigma^2 + \mu^2 l^2 + \left( \nu + \mu \frac{T}{m} \right)^2 l^{-2} \right]
\end{equation}
at time $T$. Hence, the difference between the entropy values for the measured particle and the unaffected one is 
\begin{equation}
	\Delta S_T(\mu, \nu) = \frac{1}{2} \ln \left( 1 + \frac{4}{3 \Delta a^2} \frac{3\nu^2 + 3 \nu \mu \frac{T}{m} + \mu^2 \frac{T^2}{m^2}}{ \mu^2l^2 + \left( \nu + \mu \frac{T}{m} \right)^2 l^{-2}} \right).
\end{equation} 

In the conclusion, we concern whether propagator \eqref{eq:FP_propagator} of a particle being under a position measurement is consistent with the equation \eqref{eq:propagator_master_equation} governing its time evolution. Indeed, applying the star-product formalism to equation \eqref{eq:propagator_master_equation} where $\hat{A}$ is replaced by $\hat{q}$, one thus gets \cite{Mancini1997}
\begin{equation}
	\frac{\partial \tilde{\Pi}_t}{\partial t} - \frac{\mu}{m} \frac{\partial \tilde{\Pi}_t}{\partial \nu} + f(t) \nu \frac{\partial \tilde{\Pi}_t}{\partial X} - k \nu^2 \frac{\partial^2 \tilde{\Pi}_t}{\partial X^2} = 0. 
\end{equation}
By a straightforward substitution, it can be easily verified that propagator \eqref{eq:FP_propagator} does satisfy this equation for $k = 1/\Delta a^{2}$ (note, that in order to satisfy it, one must take into account that $\Delta a^2 \sim 1/T$ \cite{Konetchnyi1993}).

\section{Conclusion}

The subject of the current study is the influence a quantum system experiences upon a continuous measurement, considered within the framework of the probability representation of quantum mechanics. 

The symplectic tomogram attributed to each isolated quantum system evolves with it, the resulting tomogram being determined by a classical propagator (or tomogram propagator). The latter, in turn, incorporates the quantum propagator of that system. We applied the representation of quantum continuous measurement through the restricted path integral to modify the classical propagator. In particular, we have introduced a partial tomogram propagator for the measurement when the result is known. This partial propagator, however, determines the tomogram density of the evolved system in the set of measurement outcomes. If one performs a non-selective measurement, when the outcome is not known, the tomogram propagator can be calculated just by integrating the partial propagators over all outcomes.

The spectral position measurement of a driven oscillator is examined as well as its particular case of a particle. Particularly, the tomogram propagator for the oscillator under a continuous position measurement is obtained. It is shown that the coordinate dependence of the tomogram of both the oscillator and the particle takes the Gaussian form. The latter results in an additional blurring of the coordinate dependence the tomogram initially had.

\bibliographystyle{ieeetr}

\bibliography{refs}

\end{document}